\documentclass[english,12pt]{article}
\usepackage{amssymb}
\usepackage{amsmath}
\usepackage{babel}
\usepackage{array}
\usepackage[dvips]{graphicx}
\usepackage{pstricks}
\usepackage{pst-node}
\input{Qcircuit}
\newcommand{\Id}{\mathbb{I}}
\newcommand{\sz}[1]{X'^{\frac{1-#1}{2}}}
\newcommand{\sx}[1]{X^{\frac{1-#1}{2}}}
\date{}
\title{Universality of Measurements on Quantum Markets}
\author{Ireneusz Paku\l a \\ Institute of Physics, University of Silesia, \\ Uniwersytecka
4, Pl 40007 Katowice, Poland \\ e-mail: ipakula@wp.pl\\ Edward W. Piotrowski\\
Institute of Mathematics, University of Bia\l ystok,\\
Lipowa 41, Pl 15424 Bia\l ystok,
Poland\\ e-mail: ep@alpha.uwb.edu.pl \\
 Jan S\l adkowski\\ Institute of Physics, University of Silesia, \\ Uniwersytecka
4, Pl 40007 Katowice, Poland \\ e-mail: sladk@us.edu.pl }
\begin{document}
\def\meter{\mbox{$\frown\hspace{-.9em}{\lower-.4ex\hbox{$_\nearrow$}}$}}
\def\circs{\mbox{$\hspace{-.3em}\bigcirc\hspace{-.69em}{\lower-.35ex\hbox{$_{\mathsf S}$}}$}}
\def\circh{\mbox{$\hspace{-.16em}\bigcirc\hspace{-.76em}{\lower-.35ex\hbox{$_{\mathsf H}$}}$}}
\maketitle
\begin{abstract}
\noindent  We reason about possible future development of quantum
game theory and its impact on information processing and  the
emerging information society. Two of the authors have recently
proposed a quantum description of financial market in terms of
quantum game theory. These "new games" cannot by themselves create
extraordinary profits or multiplication of goods, but they may
cause the  dynamism of transaction which would result in more
effective markets and capital flow into hands of the most
efficient traders. We  focus upon the problem of  universality of
measurement in quantum market games. Quantum-like approach to
market description proves to be an important theoretical tool for
investigation of computability problems in economics or game
theory even  if never implemented in real markets.
 \end{abstract}
 \vspace{5mm}

\section{Introduction}
One of the interpretations of quantum mechanics based on the
Fokker Wheeler Feynman direct interaction approach
\cite{wheeler1,wheeler2} refers to market transactions \cite{cra}.
Therefore, quantum market games \cite{PS1} have a unique bootstrap
supporting the use of quantum formalism to describe them. On the
other hand, quantum theory implies properties of players'
strategies that assuredly form new standards of market liquidity.
Quantum strategies can be identified in a nondestructive way (for
example, with a test making use of the {\em controlled-swap}
\/gate that is used in the quantum fingerprinting \cite{odcisk}),
cannot be copied nor destroyed what is guaranteed by the {\em
no-cloning}\/ and {\em no-deleting theorems}\/ \cite{hor1}. In
addition, they can be shared in a perfect, requiring no
regulations way among players-shareholders (for example in such a
way that any group of $k$ shareholders can adopt the strategy and
no smaller group of shareholders can make profit on this strategy
\cite{cleve}). Optimal management of such quantum strategies
requires an appropriate portfolio theory \cite{maur}. This should
not be regarded as a disadvantage as risk is associated even with
classical arbitrage transactions \cite{PS04} and there is a
constant need for an appropriate theory to manage the risk
associated with any activity. Currently, quantum theory is the
only one that promises this degree of perfection therefore there
is only a faint possibility that  quantum game theory might be
overvalued. Besides the number of arguments for the quantum
anthropic principle as formulated in \cite{qantpri,PS01}: even if
at earlier stages of development markets are governed by classical
laws, markets will evolve towards their quantum counterparts due
to the effectiveness of quantum mechanisms is on the in crease.
Such an evolution occurred throughout the last century, when new
technologies gained superiority over classical "common sense".
Contemporary markets undergoing a process of globalization would
intensify such evolution. Markets exploring quantum phenomena
regardless of  such details as whether its "quantumness" would
derive from instruments or human mind properties, would offer
effectiveness impossible in classical markets and therefore would
replace them sooner or later.
\section{A two-qubits dealer's strategy}
The bewildering phenomenon of quantum dense coding \cite{dense}
enables us sending two classical bit of information by exchanging
one qubit. This can be presented in the game theory setting as
follows. Suppose we intend to send the information from A to B.
Then the circuit \cite{NC}

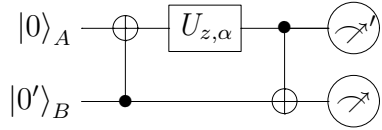
\[
\Qcircuit @C=1em @R=.7em { \lstick{\ket{0}_A\negthinspace} & \targ
& \gate{\text{\em U}_{z,\alpha}} & \ctrl{1} & \measureprim\qw
\\
\lstick{\ket{0'}_B\negthinspace} & \ctrl{-1} &\qw & \targ
&\measure\qw }
\]
\begin{equation*}\text{\em Cnot}\,(\text{\em U}_{z,\alpha}\negthinspace
\otimes I)\,\text{\em Swap\,Cnot\,Swap}\,|0\rangle_{\negthinspace
A}|0'\rangle_{\negthinspace B} \,\,=\,\,
\cos(\alpha)\,|0'\rangle_{\negthinspace A}|0\rangle_{\negthinspace
B}\,+
\end{equation*}
\begin{equation*}
\text{i}\,\sin(\alpha)\,\bigl(E_z(X)\, |0'\rangle_{\negthinspace
A}|\text{I}\rangle_{\negthinspace B}
+E_z(X')\,|\text{I}'\rangle_{\negthinspace
A}|0\rangle_{\negthinspace B} +E_z(X'')\,|\text{I}'
\rangle_{\negthinspace A}|\text{I}\rangle_{\negthinspace B}\bigr)
\end{equation*}
where
\[
X:=\sigma_x\,,\,\,X':=\sigma_z=HXH\,,\,\,X'':=\sigma_y=\text{i}XX'\,
\]
are the tactics\footnote{We call any unitary transformation that
changes agent's (player's) strategy a tactics. We follow the
notation introduced in \cite{komp}:
 $SU(2)\ni\mathcal{U}\,_{\negthinspace z,\alpha}=\,
\mathrm{e}^{\text{i}\alpha\overrightarrow{\sigma}\cdot
E_z(\overrightarrow{\sigma})}\,=\,I\cos\alpha
+\text{i}\,\overrightarrow{\sigma} \negthinspace\cdot\negthinspace
E_z(\overrightarrow{\sigma})\,\sin\alpha\, $, where the vector
$E_z(\overrightarrow{\sigma})= \frac{\langle
z|\overrightarrow{\sigma}|z\rangle} {\langle z|z\rangle} $
represents the expectation value of the vector of Pauli matrices
$\overrightarrow{\sigma}:=(\sigma_1,\sigma_2,\sigma_3)$ for a
given strategy $|z\rangle$. The family
$\{|z\rangle\}$,$z\negthinspace\in\negthinspace\overline{\mathbb{C}}$
of complex vectors (states)
$|z\rangle:=|0\rangle+z\,|\text{I}\rangle$
($|\negthinspace\pm\negthinspace\infty\rangle:=|\text{I}\rangle$)
represents all trader's strategies in the linear subspace spanned
by the vectors $|0\rangle$ and $|\text{I}\rangle$.} given by the
Pauli matrices, describes such a process. The representation of
the tactics $\text{\em U}_{z,\alpha}$ in terms of the final
strategy is utterly secure because the owner of the pair of qubits
$A$ and $B$ can keep the information distinguishing these two
qubist (one classical bit) secret; without this information the
interception of the pair of qubits $A$ and $B$ is insufficient for
identification of the tactics $\text{\em U}_{z,\alpha}$. Such a
method has previously been applied by Wiesner to construct quantum
counterfeit-proof banknotes \cite{Wie}. By adopting the tactics
$\text{\em U}_{z,\alpha}$ that corresponds to one of four pairwise
maximally distant pairs of antipodal points of the sphere $S_3$
\footnote{ $ U = a_0 I + \text{i} \sum_k a_k \sigma_k $, where $a_0 = cos
\alpha$, $a_k = n_k sin \alpha$ and $ \sum_\mu (a_\mu)^2 =1$. The
corresponding tactics are $ \pm I \pm \sigma_k $, where the
antipodal points  have different signs but represent equivalent
tactics.} the owner of qubit $A$ is able to send two classical
bits to the owner of qubit $B$ while sending only one qubit.
According to the analysis given in Ref\mbox{.} \cite{PS6}, the measurable
qubits $B$ and $A$ can be interpreted as market polarizations of
their owner (if $|0\rangle$ -- supply and if $|1\rangle$ -- demand)
and therefore his/her inclination to buy at low or high prices
what can easily be seen if we replace the meters with the
controlled-Hadamard gates with control qubit $B$. In order to
 connect unequivocally any of the three conjugated bases \cite{Wie}
(or mutually unbiased \cite{Woott}) with one of their three
possible market functions (eigenvectors (fixed points) of $X$ with
supply inclination, eigenvectors of $X'$ with demand inclination
and eigenvectors of $X''$ with polarization) we should transform
the strategy $B$ (after the controlled-Hadamard gate!) with the
involutive tactics $G$ :
\[
G:=\tfrac{1}{\sqrt{2}}(X'+X'')\,.
\]
that transforms eigenvectors of  $X'$ into eigenvectors of $X''$.
The consideration of the third conjugated basis is necessary to
guarantee the security of the information a la Wiesner' banknotes
(the information about the respective price carried by qubit $A$
uses two conjugated bases -- sets of fixed points of  tactics $X$
and $X'$).
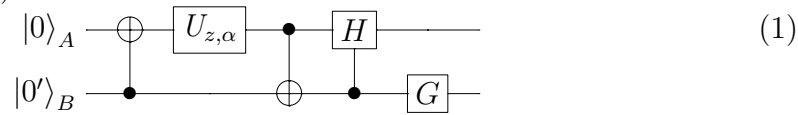
\begin{equation}
\Qcircuit @C=1em @R=.7em { \lstick{\ket{0}_A\negthinspace} & \targ
& \gate{\text{\em U}_{z,\alpha}} & \ctrl{1} & \gate{H} &\qw&\qw
\\
\lstick{\ket{0'}_B\negthinspace} & \ctrl{-1} &\qw & \targ
&\ctrl{-1}&\gate{G}&\qw } \label{chargate}
\end{equation}
If there is no restriction on the tactics $\text{\em
U}_{z,\alpha}$ (we can even consider the whole two-dimensional
Hilbert space as the set of allowed strategies) the agent is able
to play more effectively by adopting superpositions of previously
allowed strategies. We have already considered alliances
(implemented as gates between qubit strategies); they are
universal. Therefore, if alliances are allowed tactics quantum
market is equivalent to quantum computer! Obviously, quantum
markets cam have various different properties -- the polarization
qubit is redundant in two-sided auctions but  in bargaining games
\cite{PS6} another qubit  is necessary to distinguish the agents
who are  bidding. Much more additional qubits are necessary if the
corresponding supply and demand curves are continuous (floating
point precision) -- one qubit for each binary digit of the
logarithm of price). However, these are theoretically unimportant
details -- all such forms of quantum markets can be implemented
with the use of elementary market measurements alone what follows
from the analysis by Nielsen, Raussendorf and Briegel, and Perdrix
 and Jorrand \cite{N}-\cite{jorrand1}. The rest of the paper is devoted to this problem.
 Note that such a dominant role of market measurements suggests that quantum market
 may be free from psychological factors, such as phobia, intention,
 irrationality and so forth. Besides, all classical models of markets are
 limiting cases of quantum models in an analogous way to the
 transition from quantum to classical mechanics. The variety of quantum games forming evolving
 towards better effectiveness information oriented quantum markets supplements the idea of quantum
 darwinism put forward by \.Zurek \cite{zur-dar}. Therefore,
 the popular but never proved hypothesis of  "humanism of markets" is only an
 illusion. Nevertheless, we should warn the reader that quantum dynamics may
 result in effects that even philosopher would not dare to dream of \cite{renninger}.

\section{Measurements of tactics}
A measurement of tactics consists in determination of the strategy
or, more precisely, discoveries which of its fixed points we have
to deal with. If the tactics being measured changes the
corresponding strategy, then the non-demolition measurement
reduces the strategy to one of its fixed points and the respective
transition amplitudes are given  by coordinates of the strategy in
the fixed point basis (Born rule). As we will show,  measurements
of the tactics $X$, $G$ and $X\otimes X'$ suffice to implement
quantum market games. According to the {\it Qcircuit.tex}\/
standard macros \cite{eastin}, we will denote the corresponding
measuring gates as (rounded off shape is used to distinguish
measuring gates):
\begin{equation} \Qcircuit @C=1em @R=.7em {
 &\bgate{X}&\qw
} \,\,,\,\, \Qcircuit @C=1em @R=.7em {
 &\bgate{G}&\qw
} \,\,,\,\, \lower-.9em\hbox{\Qcircuit @C=1em @R=.7em{
&\bmultigate{1}{X\negthinspace\otimes\negthinspace X'}&\qw\\
&\ghost{X\negthinspace\otimes\negthinspace X'}&\qw }}\,\,.
\label{uniwersalne}
\end{equation}
Note that measurement of the tactics
$X\negthinspace\otimes\negthinspace X'$ provides us with
information whether the two strategies  agree or disagree on the
price but reveals no information on the level of the price in
question. To get information about the prices we have to measure
$X\otimes I$ and $I\otimes X'$ respectively. Note that the
measurement of $X'$ can be implicitly accomplished   by
measurement of $X$ and subsequently $X\otimes X'$. This is shown
graphically by
\[
\Qcircuit @C=1em @R=.7em{
{\scriptstyle(}&\bgate{X}&\bmultigate{1}{{X\negthinspace\otimes\negthinspace
X'}}&
\qw {\scriptstyle)}\\
&\qw&\ghost{X\negthinspace\otimes\negthinspace X'}&\qw }\,\,
\lower.7em\hbox{\,\,\,\,$\Longrightarrow$\,\,\,} \lower.4em\hbox{
\,\,\Qcircuit @C=1em @R=.7em { &\bgate{X'}&\qw }\,\, ,}
\]
where the parentheses are used to denote auxiliary qubits. In the
following paragraphs we will analyze q-circuits with various
number of auxiliary qubits that would allow for implementation of
tactics via measurement only -- the approach proposed by S. Perdix
\cite{Per}.
\section{Universality of measurements: implementing tactics via
measurements}

Teleportation and measurement form  surprisingly powerful tools in
implementation of tactics. The method used by Perdrix and Jorrand
\cite{Per,jorrand1} to analyse the problem of universality in
quantum computation can be easily adopted to the situation we are
considering.  Following Ref. \cite{Per}, we  begin by showing how
a strategy encoded in one qubit can be transferred to another
(from the upper one to the lower one in the figure below) and how
it changes with a sequence of tactics $\sigma H$, where
 $\sigma$ is one of the
Pauli matrices (including the identity matrix):
\begin{equation}
\Qcircuit @C=1em @R=.7em{
&\qw&\bmultigate{1}{{X\negthinspace\otimes\negthinspace X'}}&
\bgate{X'}&\qw {\scriptstyle)}\\
{\scriptstyle(}&\bgate{X}&\ghost{X\negthinspace\otimes\negthinspace
X'}&\qw&\qw }\,\, \lower.7em\hbox{\,\,\,\,$\Longrightarrow$\,\,\,}
\lower.4em\hbox{ \,\,\Qcircuit @C=1em @R=.7em { &\gate{\sigma
H}&\qw }\,\,.} \label{hgate}
\end{equation}

Assuming that the input qubit is in the state:
\[
\ket{\psi}=\alpha\ket{0}+\beta\ket{1},
\]
after measuring $\Id\otimes X$ (with classical outcome  $j=\pm 1$)
we obtain:
\[
\ket{\psi_1}=\ket{\psi}\otimes\sz{j}\frac{\ket{0}+\ket{1}}{\sqrt{2}}=
\frac{1}{\sqrt{2}} (\Id \otimes\sz{j})
(\alpha\ket{00}+\alpha\ket{01}+\beta\ket{10}+\beta\ket{11}).
\]
Measurement of $X\otimes X'$ with outcome $k=\pm 1$ sets our qubits in state:
\[
\ket{\psi_2}=\frac{1}{\sqrt{2}}(\Id\otimes\sx{k}\sz{j})\left[(\alpha+\beta)(\ket{00}+\ket{10})
+(\alpha-\beta)(\ket{01}-\ket{11})\right].
\]
The final measurement $X'\otimes\Id$ with outcome $l=\pm 1$ gives
us the final state:
\begin{eqnarray*}
&\ket{\psi_3}
=\left[\sx{l}\otimes\sx{k}\sz{j}H\sx{l}\right]\left[\ket{0}\otimes (\alpha\ket{0}+\beta\ket{1})\right]=&\\
&=\sx{l}\ket{0}\otimes \sx{k}\sz{j\cdot l}H\ket{\psi},&
\end{eqnarray*}
and the equivalence of the circuits above is proved.

Thus, the strategy encoded in the upper state  is transferred from
 the lower qubit and changed with the tactics $\sigma H$, where
$\sigma= \sx{k}\sz{j\cdot l}$. It is evident that the same tactics
is adopted when we switch the supply measurements with the demand
ones ($X\negthinspace\leftrightarrow\negthinspace X'$). Simple
calculation shows that the composite tactics $H\sigma H$ and
$\sigma _{i}\sigma _{k}$ reduce to some Pauli (matrix) tactics.
Therefore, an even sequence of tactics $(\ref{hgate})$ can be
perceived as the Markov process over vertices of the graph 
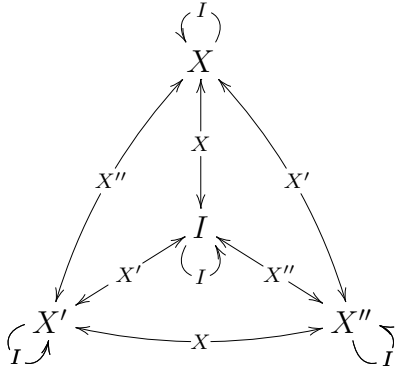
\[ \xy
(0,1)*+{I}="a",(0,23.1)*+{X}="b",(-20,-11.55)*+{X'}="c",(20,-11.55)*+{X''}="d"
\ar@(ld,rd)|{\,I\,}"a";"a" \ar@(ur,ul)|{\;I\,}"b";"b"
\ar@(l,d)|{\;I\,} "c";"c" \ar@(d,r)|{\vphantom{^I}\;I\,}"d";"d"
\ar@{<->}@/^1.5ex/|{\,\,\vphantom{\sum^I}X'}"b";"d"
\ar@{<->}@/_1.5ex/|{\vphantom{\sum^I}X''\negthinspace\negthinspace}"b";"c"
\ar@{<->}|{\vphantom{\sum^I}X}"b";"a" \ar@(l,d)|{\;I\,}"c";"c"
\ar@{<->}|{\,\,\vphantom{\sum^I}X'}"c";"a"
\ar@(d,r)|{\vphantom{^I}\;I\,}"d";"d"
\ar@{<->}|{\vphantom{\sum^I}X''\negthinspace\negthinspace}"d";"a"
\ar@{<->}@/^1.5ex/|{\vphantom{\sum^I}X}"d";"c"
\endxy
\]
It follows that  any Pauli tactics can be implemented  as an even
number of tactics-measurements $(\ref{hgate})$ by identifying it
with some final vertex of random walk on this graph. Although the
probability of drawing out the final vertex at the first step is
$\tfrac{1}{4}$, the probability of staying in the "labyrinth"
exponentially decreases to zero. Having a method of implementation
of Pauli tactics, allows us to modify the tactics $(\ref{hgate})$
so that to implement the tactics $H$ -- the fundamental operation
of switching the supply representation with the demand
representation. It can be also applied to measure compliance with
tactics representing the same side of the market (direct
measurement is not possible because  the agents cannot make the
deal):
\[
\Qcircuit @C=1em @R=.7em{
&\bmultigate{1}{X\negthinspace\otimes\negthinspace X}&\qw\\
&\ghost{X\negthinspace\otimes\negthinspace X}&\qw }
\lower.6em\hbox{\,\,\,$:=$\,\,\,} \Qcircuit @C=1em @R=.7em{
&\qw&\bmultigate{1}{X\negthinspace\otimes\negthinspace X'}&\qw&\qw\\
&\gate{H}&\ghost{X\negthinspace\otimes\negthinspace
X'}&\gate{H}&\qw }\,\,\lower.6em\hbox{\,,}
\]
\[
\Qcircuit @C=1em @R=.7em{
&\bmultigate{1}{X'\negthinspace\otimes\negthinspace X'}&\qw\\
&\ghost{X'\negthinspace\otimes\negthinspace X'}&\qw }
\lower.6em\hbox{\,\,\,$:=$\,\,\,} \Qcircuit @C=1em @R=.7em{
&\gate{H}&\bmultigate{1}{X\negthinspace\otimes\negthinspace X'}&\gate{H}&\qw\\
&\qw&\ghost{X\negthinspace\otimes\negthinspace X'}&\qw&\qw
}\lower1em\hbox{\,\,\,.}
\]
In addition, this would allow for interpretation via measurement
of random Pauli tactics $\sigma$ because of the involutiveness of
$H$ the gate $(\ref{hgate})$ can be transformed to
\begin{equation}
\Qcircuit @C=1em @R=.7em{
&\gate{H}&\bmultigate{1}{{X\negthinspace\otimes\negthinspace X'}}&
\bgate{X'}&\qw {\scriptstyle)}\\
{\scriptstyle(}&\bgate{X}&\ghost{X\negthinspace\otimes\negthinspace
X'}&\qw&\qw }\lower.8em\hbox{\,\,\,\,$=$\,\,\,\,} \Qcircuit @C=1em
@R=.7em{ &\qw&\bmultigate{1}{{X'\negthinspace\otimes\negthinspace
X'}}&
\bgate{X}&\qw {\scriptstyle)}\\
{\scriptstyle(}&\bgate{X}&\ghost{X'\negthinspace\otimes\negthinspace
X'}&\qw&\qw } \lower.8em\hbox{\,\,\,\,$=$} \label{sigmagate}
\end{equation}
\[
\Qcircuit @C=1em @R=.7em{
&\qw&\bmultigate{1}{{X\negthinspace\otimes\negthinspace X}}&
\bgate{X'}&\qw {\scriptstyle)}\\
{\scriptstyle(}&\bgate{X'}&\ghost{X\negthinspace\otimes\negthinspace
X}&\qw&\qw } \,\, \lower.7em\hbox{\,\,\,\,$\Longrightarrow$\,\,\,}
\lower.4em\hbox{ \,\,\Qcircuit @C=1em @R=.7em { &\gate{\sigma}&\qw
}\,\,.}
\]
The gate $(\ref{sigmagate})$ can be used to implement the
phase-shift tactics:
\[T:=\begin{pmatrix}
  1& 0 \\
  0& \tfrac{1+\text{i}}{\sqrt{2}}
\end{pmatrix}\,.
\]
$T$ commutes with  $X'$, hence:
\[
\lower.4em\hbox{ \,\,\Qcircuit @C=1em @R=.7em { &\gate{\sigma
T}&\qw }} \lower.7em\hbox{\,\,\,\,$\Longleftarrow$\,\,\,}
\Qcircuit @C=1em @R=.7em{
&\qw&\bmultigate{1}{{X'\negthinspace\otimes\negthinspace X'}}&
\bgate{T^{-1}X\,T}&\qw {\scriptstyle)}\\
{\scriptstyle(}&\bgate{X}&\ghost{X'\negthinspace\otimes\negthinspace
X'}&\qw&\qw } \lower1em\hbox{\,\,\,.}
\]
Elementary  calculation demonstrates that
$T^{-1}XT\negthinspace=\negthinspace\tfrac{X-X''}{\sqrt{2}}$ and
$H\tfrac{X-X''}{\sqrt{2}}H\negthinspace=\negthinspace G$,
therefore:
\[\Qcircuit @C=1em @R=.7em{
&\gate{H}&\bmultigate{1}{{X\negthinspace\otimes\negthinspace X'}}&
\bgate{G}&\qw {\scriptstyle)}\\
{\scriptstyle(}&\bgate{X}&\ghost{X\negthinspace\otimes\negthinspace
X'}&\qw&\qw } \,\,
\lower.7em\hbox{\,\,\,\,$\Longrightarrow$\,\,\,} \lower.4em\hbox{
\,\,\Qcircuit @C=1em @R=.7em { &\gate{\sigma T}&\qw }}
\lower1em\hbox{\,\,\,.}
\]
We have seen earlier that it is possible to remove the superfluous
Pauli operators, cf. $(\ref{sigmagate})$. To end the proof of
universality of the set of gates $(\ref{uniwersalne})$ we have to
show how to implement the alliance $Cnot$ (note that
$\{H,T,Cnot\}$ a set of universal gates \cite{bar}). This gate can
be implemented as the circuit (as before, the gate is constructed
up to a Pauli tactics) \cite{Per}:
\begin{equation}
\Qcircuit @C=1em @R=.7em{
&\qw&\qw&\bmultigate{1}{X'\negthinspace\otimes\negthinspace X}&\qw&\qw\\
{\scriptstyle(}&\bgate{X}&\bmultigate{1}{X'\negthinspace\otimes\negthinspace
X}&
\ghost{{X'\negthinspace\otimes\negthinspace X}}&\bgate{X'}&\qw{\scriptstyle)} \\
&\qw&\ghost{X'\negthinspace\otimes\negthinspace X}&\qw&\qw&\qw
}\,\, \lower1.6em\hbox{\,\,\,\,\,\,$\Longrightarrow$\,\,\,}
\lower.5em\hbox{ \,\,\Qcircuit @C=1em @R=.7em {
&\ctrl{1}&\gate{\sigma_a}&\qw\\
&\targ&\gate{\sigma_b}&\qw
}\,\,\lower1em\hbox{\,\,\,.}}\label{grafjs}
\end{equation}
The explicit calculations are as follows\footnote{The tactics $H$
transforms the demand picture to the supply picture
($X\negthinspace\leftrightarrow\negthinspace X'$),  what results
in a switch from control qubit to the controlled qubit.}:

Let us assume the input qubits are the first and the second, in
state:
\[
\ket{\psi}=\alpha\ket{00}+\beta\ket{01}+\gamma\ket{10}+\delta\ket{11}
\]
with the third used as an auxiliary one.
States $\ket{\pm}$ are defined as follows:
\[
\ket{\pm}=\frac{1}{\sqrt{2}}(\ket{0}\pm\ket{1}).
\]
The first measurement $\Id\otimes\Id\otimes X$ gives us the state
below, depending on classical outcome $j=\pm 1$\footnote{Note that
 the second and the third qubit appear in reversed order in Fig.$(\ref{grafjs})$ }:
\begin{eqnarray*}
&\ket{\psi_1}=\ket{\psi}\otimes\sz{j}\ket{+}=&\\
&(\Id\otimes\Id\otimes\sz{j})(\alpha\ket{00+}+
\beta\ket{01+}+\gamma\ket{10+}+\delta\ket{11+}).&
\end{eqnarray*}
After $\Id\otimes X \otimes X'$ with outcome $k=\pm 1$ we obtain:
\begin{eqnarray*}
&\ket{\psi_2}=\frac{1}{2}\left[\Id\otimes\sz{k}\otimes\sz{j}\right]\times&\\
&\times\lbrace (\alpha+\beta)(\ket{000}+\ket{010})+(\alpha-\beta)(\ket{001}-\ket{011})+&\\
&+(\gamma+\delta)(\ket{100}+\ket{110})+(\gamma-\delta)(\ket{101}-\ket{111})\rbrace=&\\
&=\frac{1}{\sqrt{2}}\left[\Id\otimes\sz{k}\otimes\sz{j}\right]
\lbrace(\alpha+\beta)\ket{0+0}+(\alpha-\beta)\ket{0-1}+&\\
&+(\gamma+\delta)\ket{1+0}+(\gamma-\delta)\ket{1-1}\rbrace.&
\end{eqnarray*}
Next measurement $X'\otimes\Id\otimes X$ with outcome $l=\pm 1$ sets our qubits in state:
\begin{eqnarray*}
&\ket{\psi_3}=\left[\Id\otimes\sz{k}\sx{l\cdot j}\otimes\sz{l}\right]
\left[\alpha\ket{00+}+\beta\ket{01+}+\delta\ket{10-}+\gamma\ket{11-}\right].&
\end{eqnarray*}
After the final measurement of $\Id\otimes\Id\otimes X'$ with
eigenvalues $m=\pm 1$ we get:
\begin{eqnarray*}
&\ket{\psi_3}=\left[\sz{k}\otimes\sx{l\cdot j}\otimes\sz{l}\right]
\left[\alpha\ket{00+}+\beta\ket{01+}+\delta\ket{10-}+\gamma\ket{11-}\right].&
\end{eqnarray*}
After the final measurement of $\Id\otimes\Id\otimes X'$ with
eigenvalues $m=\pm 1$ we get:
\begin{eqnarray*}
&\ket{\psi_4}=\left[\sz{m\cdot k}\otimes\sx{l\cdot j}\otimes\sx{m}\right]\times&\\
&\times\left[(\alpha\ket{00}+\beta\ket{01}+\delta\ket{10}+\gamma\ket{11})\otimes
\ket{0}\right]=&\\
&=\left[\sz{m\cdot k}\otimes\sx{l\cdot j}\otimes\sx{m}\right]
\left[CNot\ket{\psi}\otimes\ket{0}\right],&
\end{eqnarray*}
thus proving the above circuit equivalence.

The measurement of the tactics $G$ conducted within the quantum
market game frame of reference causes  interpretative problems
that can be resolved if we replace the measurement of $G$ with
$controlled \ H$ gate (cf $(\ref{chargate})$) and the measurement
of entanglement for another pair of conjugated bases
$X'\negthinspace\otimes\negthinspace X''$ in the set of universal
primitives. Owing to the fact that $G=HGHGH$, the measurement of
$G$ can be implemented in the following way \cite{NC}:
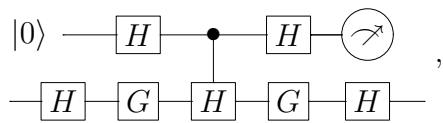
\[
\Qcircuit @C=1em @R=.7em{
&\lstick{\ket{0}}&\gate{H}&\ctrl{1}&\gate{H}&\measure\qw&\\
&\gate{H}&\gate{G}&\gate{H}&\gate{G}&\gate{H}&\qw
}\,\lower.6em\hbox{\,,}
\]
where the tactics $G$ is obtained from $(\ref{hgate})$ by the
cyclic replacement $X\negthinspace\rightarrow\negthinspace X'$,
$X'\negthinspace\rightarrow\negthinspace X''$ i
$X''\negthinspace\rightarrow\negthinspace X$:
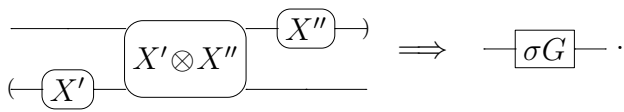
\begin{equation*}
\Qcircuit @C=1em @R=.7em{
&\qw&\bmultigate{1}{{X'\negthinspace\otimes\negthinspace X''}}&
\bgate{X''}&\qw {\scriptstyle)}\\
{\scriptstyle(}&\bgate{X'}&\ghost{X'\negthinspace\otimes\negthinspace
X''}&\qw&\qw }\,\,
\lower.7em\hbox{\,\,\,\,$\Longrightarrow$\,\,\,} \lower.4em\hbox{
\,\,\Qcircuit @C=1em @R=.7em { &\gate{\sigma G}&\qw }\,\,.}
\end{equation*}
In fact,  the universality property has any set of primitive that
contains the $controlled\ H$ gate and measurements $X^k,
X^p\negthinspace\otimes\negthinspace X^q,
X^r\negthinspace\otimes\negthinspace X^s$, where
$p\negthinspace\neq\negthinspace q$,
$r\negthinspace\neq\negthinspace s$ and
$p\negthinspace\neq\negthinspace r$ -- this can be easily checked
\cite{Per}. It follows that to implement a quantum
market\footnote{In fact, any finite-dimensional quantum
computational system can be implemented in that way \cite{Per}.}
it suffices to have, beside possibility of measuring
strategy-qubits and control of the supply-demand context, a direct
method measuring entanglement of a pair of qubits in conjugated
bases.
\section{Quantum Intelligence \`a la  20 questions }
Let us recall the anecdote popularized by John Archibald Wheeler
\cite{d-b}. The plot concerns the game of 20 questions: the player
has to guess an unknown word by asking up to 20 questions (the
answers could be only yes or no and  are always true). In the
version presented by Wheeler, the answers are given by a "quantum
agent" who attempts to asign the task the highest level of
difficulty without breaking the rules. In the light of the
previous discussion, any quantum algorithm (including classical
algorithms as a special cases) can be implemented  as a sequence
of appropriately constructed questions-measurements. The results
of the measurements (ie answers) that are not satisfactory cause
further "interrogation" about selected elementary ingredients of
the reality (qubits). If Quantum Intelligence (QI) is perceived in
such a way (as quantum game) then it can be simulated  by a
deterministic automaton that follows a chain of test bits built on
a quantum tenor \cite{Deu98}. The automaton completes the chain
with afore prepared additional questions at any time that an
unexpected answer is produced (cf the idea of quantum darwinism
\cite{zur-dar}). Although the results of the test will be random
(and actually meaningless -- they are instrumental), the kind and
the topology of tests that examine various layers multi-qubit
reality and the working scheme of the automaton are fixed prior to
the test. The remarkability of performance of such an automaton in
a game against Nature is by the final measurement that could
reveal knowledge that is out of reach of classical information
processing, cf the already known Grover and Shor quantum
algorithms and the Elitzur-Vaidman bomb tester. Needless to say,
such an implementation of a game against quantum Nature leaves
some room for perfection. The tactics $CNot$ and  $H$ belong to
the normalizer of the $n$-qubit Pauli group  $G_n$ \cite{NC},
hence their adoption allows to restrict oneself to single
corrections of "errors" made by Nature that precede the final
measurement.  It is worth noting that a variant of implementation
of the tactics  $T$ makes it possible to postpone the correction
provided the respective measurements methods concern the current
state of the cumulated errors \cite{jorrand2}. Therefore in this
setting of the game some answers given by Nature, though being
instrumental, have a significance because of the influence of the
following tests. There is no need for the final error correction
-- a modification of the measuring method is sufficient. In that
way the course of game is fast and the length of the game is not a
random variable. This example shows that in some sense the
randomness  in game against quantum Nature can result from
awkwardness of agents and erroneous misinterpretation of answers
that are purely instrumental.

 {\bf Acknowledgements} This paper has been supported
by the {\bf Polish Ministry of Scientific Research and Information
Technology} under the (solicited) grant No {\bf
PBZ-MIN-008/P03/2003}.

\end{document}